\documentclass[pra,print,showpacs,superscriptaddress,twocolumn]{revtex4}
\usepackage{amssymb}
\usepackage{graphicx}
\usepackage{dcolumn}
\usepackage{bm}
\usepackage{amsmath}

\begin{document}

\title{Spin-orbit-induced bound state and molecular signature of the
degenerate Fermi gas in a narrow Feshbach resonance}
\author{Kuang Zhang}
\affiliation{State Key Laboratory of Quantum Optics and Quantum Optics Devices, Institute
of Laser spectroscopy, Shanxi University, Taiyuan 030006, People's Republic
of China}
\author{Gang Chen}
\thanks{Corresponding author: chengang971@163.com}
\affiliation{State Key Laboratory of Quantum Optics and Quantum Optics Devices, Institute
of Laser spectroscopy, Shanxi University, Taiyuan 030006, People's Republic
of China}
\author{Suotang Jia}
\affiliation{State Key Laboratory of Quantum Optics and Quantum Optics Devices, Institute
of Laser spectroscopy, Shanxi University, Taiyuan 030006, People's Republic
of China}

\begin{abstract}
In this paper we explore the spin-orbit-induced bound state and molecular
signature of the degenerate Fermi gas in a narrow Feshbach resonance based
on a generalized two-channel model. Without the atom-atom interactions, only
one bound state can be found even if spin-orbit coupling exists. Moreover,
the corresponding bound-state energy depends strongly on the strength of
spin-orbit coupling, but is influenced slightly by its type. In addition, we
find that when increasing the strength of spin-orbit coupling, the critical
point at which the molecular fraction vanishes shifts from zero to the
negative detuning. In the weak spin-orbit coupling, this shifting is
proportional to the square of its strength. Finally, we also show that the
molecular fraction can be well controlled by spin-orbit coupling.
\end{abstract}

\maketitle

\section{Introduction}

Recently, the investigation of spin-orbit (SO) coupling in neutral atoms has
attracted much attentions \cite{JR}. In particular, a one-dimensional (1D)
equal Rashba and Dresselhaus SO coupling has been first realized in the
ultracold $^{87}$Rb atoms by a couple of Raman lasers \cite{Lin}. By
applying the same laser technique, this 1D SO coupling has been also
achieved experimentally in the degenerate Fermi gas with $^{40}$K \cite{PW}
and $^{6}$Li \cite{LWC}. Theoretical investigations have been revealed that
in the presence of SO coupling, the degenerate Fermi gas can exhibit the
interesting physics in both three \cite%
{JPV,GM,YZQ,HH,JPV1,Iskin,LJ,WY1,LD,Li,KS,KZ,RL,ZhangP1,Peng1,HLY1,LD1} and
lower dimensions \cite{Chen,MG,HLY2,JZ,WY2,XYang,ZJN,FW}. For example, by
increasing the strength of SO coupling, the density of state at the Fermi
surface is increased, and the Cooper paring gap can be thus enhanced
significantly \cite{GM,YZQ,HH}. More importantly, this system may be changed
from the Bardeen-Cooper-Schrieffer (BCS) superfluid to the Bose-Einstein
condensate (BEC) with a new two-body bound state called Rashbon \cite{JPV,HH}%
. When an effective Zeeman field is applied, the 2D degenerate Fermi gas
with the Rashba SO coupling exhibits an exotic topological superfluid
supporting the Majorana fermions \cite{MG}, which is the heart for realizing
the topological quantum computing \cite{CN}. Recently, a universal midgap
bound state in the topological superfluid has been predicted \cite{HH1}.

To illustrate the SO-driven fundamental physics, a generalized one-channel
model has been introduced in all previous considerations \cite%
{JPV,GM,YZQ,HH,JPV1,Iskin,LJ,WY1,LD,Li,KS,KZ,RL,ZhangP1,Peng1,HLY1,LD1,Chen,MG,HLY2,JZ,WY2,XYang,ZJN,FW}%
. In this one-channel model, only the atoms tuned via the Feshbash-resonant
technique are taken into account. However, it is valid for the \emph{broad}
Feshbash-resonant regime with $\Gamma \gg 1$ \cite{DES}, where the
dimensionless parameter is defined as $\Gamma =\sqrt{32m\mu _{\text{B}}a_{%
\text{bg}}^{2}B_{\text{w}}^{2}/(\hbar ^{2}\pi E_{\text{F}})}$, $\mu _{\text{B%
}}$ is the Bohr magneton, $m$ is the atom mass, $a_{\text{bg}}$ is the
background $s$-wave scattering strength, $B_{\text{w}}$ is the resonant
width and $E_{\text{F}}$ is the Fermi energy. In fact, to get a more
realistic and complete description of the degenerate Fermi gas, especially
in the \emph{narrow} Feshbash-resonant limit with $\Gamma \ll 1$, we must
introduce a two-channel model \cite{MH,ETI,YO,RDU,Peng2}, which includes
both the atoms in the open channel and the molecules in the closed channel.
Moreover, in the narrow Feshbash-resonant regime, some fundamental
properties can be observed experimentally by detecting the striking
molecular signature \cite{GBP}, additional to measuring the superfluid
pairing gap applied usually in the one-channel model \cite{Chin2004,Shin,STG}%
. More importantly, new quantum phase transitions have been predicted \cite%
{YN,ZS}, attributed to the existence of extra U(1) symmetry for the
molecular field. On the experimental side, the degenerate Fermi gas in the
narrow Feshbach-resonant regime has been also reported successfully in $^{6}$%
Li \cite{ELH} and the Fermi-Fermi mixture of $^{6}$Li and $^{40}$K \cite{LC}%
. Thus, it is crucially important to explore the SO-induced exotic physics
in this regime \cite{JXC}.

In the present paper we investigate the SO-induced bound state and molecular
signature of the degenerate Fermi gas in the narrow Feshbach resonance. The
main results are given as follows. (i) Without the atom-atom interactions,
only one bound state can be found even if SO coupling exists. Moreover, the
corresponding bound-state energy depends strongly on the strength of SO
coupling, but is influenced slightly on its type. (ii) With the increasing
of the strength of SO coupling, the critical point at which the molecular
fraction vanishes shifts from zero to the negative detuning. In the weak SO
coupling, this shifting is proportional to the square of its strength. (iii)
Finally, we also show that the molecular fraction can be well controlled by
SO coupling. We believe that in experiments it is a good signature to detect
the SO-induced physics.

\section{Model and Hamiltonian}

For the SO-driven two-channel model, the total Hamiltonian can be written
formally as
\begin{equation}
H=H_{\text{F}}+H_{\text{M}}+H_{\text{I}}+H_{\text{S}}.  \label{TOTAL H}
\end{equation}%
In Hamiltonian (\ref{TOTAL H}),
\begin{equation}
H_{\text{F}}=\sum_{\mathbf{k}\sigma }\epsilon _{\mathbf{k}}C_{\mathbf{k}%
\sigma }^{\dagger }C_{\mathbf{k}\sigma }  \label{HF}
\end{equation}%
is the atom Hamiltonian, where $C_{\mathbf{k}\sigma }^{\dagger }$ is the
creation operator for a atom with the momentum $\mathbf{k}$ and $\sigma
=\uparrow ,\downarrow $, and $\epsilon _{\mathbf{k}}=k^{2}/(2m)$\textbf{\ }%
is the kinetic energy of the atom.
\begin{equation}
H_{\text{M}}=\sum_{\mathbf{q}}(\epsilon _{\mathbf{q}}+\delta _{0})b_{\mathbf{%
q}}^{\dagger }b_{\mathbf{q}}  \label{HM}
\end{equation}%
is the molecular Hamiltonian, where $b_{\mathbf{q}}^{\dagger }$ is the
creation operator of a molecule with the momentum $\mathbf{q}$, $\epsilon
_{q}=\hbar ^{2}q^{2}/(2M)$ with $M=2m$ is the kinetic energy of the
molecule, and $\delta _{0}$ is the bare detuning determined by the
Feshbash-resonant position $\delta $ via a relation
\begin{equation}
\delta _{0}=\delta +g^{2}\sum_{\mathbf{k}}\frac{1}{2\epsilon _{\mathbf{k}}}.
\label{REA}
\end{equation}%
Without SO coupling, the system has the BCS superfluid in the positive
detuning ($\delta >0$), and enters into the BEC regime in the negative%
\textbf{\ }detuning ($\delta <0$)\textbf{\ }\cite{MH,ETI,YO,RDU,Peng2}%
\textbf{.} At lower energy, the position is given approximately by $\delta
\simeq 2\mu _{\text{B}}(B-B_{0})$, where $B_{0}$ is the magnetic field at
which the resonance is at zero energy, and $B$ is the tunable magnetic field
\cite{Chin}. The atom-molecule interconversion term is governed by the
following Hamiltonian
\begin{equation}
H_{\text{I}}=g\sum_{\mathbf{qkk}^{\prime }}b_{\mathbf{q}}^{\dagger }C_{-%
\mathbf{k}+\frac{\mathbf{q}}{2}\downarrow }C_{\mathbf{k}+\frac{\mathbf{q}}{2}%
\uparrow }+C_{\mathbf{k}^{\prime }+\frac{\mathbf{q}}{2}\uparrow }^{\dagger
}C_{-\mathbf{k}^{\prime }+\frac{\mathbf{q}}{2}\downarrow }^{\dagger }b_{%
\mathbf{q}},  \label{HI}
\end{equation}%
where $g$ is the coupling constant that measures the amplitude of the decay
of the molecule in the closed channel into a pair of the open-channel atoms.%
\textbf{\ }Finally, the SO coupling is chosen as a generalized Rashba and
Dresselhaus type. The corresponding Hamiltonian is given by
\begin{equation}
H_{\text{S}}=\alpha \sum_{\mathbf{k}}[(k_{y}+i\lambda k_{x})C_{\mathbf{k}%
\uparrow }^{\dagger }C_{\mathbf{k}\downarrow }+(k_{y}-i\lambda k_{x})C_{%
\mathbf{k}\downarrow }^{\dagger }C_{\mathbf{k}\uparrow }]  \label{HS}
\end{equation}%
with $\alpha =(\alpha _{R}+\alpha _{D})$ and $\lambda =(\alpha _{R}-\alpha
_{D})/(\alpha _{R}+\alpha _{D})$, where $\alpha _{R}$ and $\alpha _{D}$ are
the SO coupling strengths for the Rashba and Dresselhaus types,
respectively. Clearly, $\alpha $ is the generalized strength of SO coupling
and the dimensionless parameter $\lambda \ $reflects the competition between
these different types of SO coupling. For example, for $\lambda =1$ ($\alpha
_{D}=0$), the 2D Rashba SO coupling can be found. Whereas, for $\lambda =0$ (%
$\alpha _{D}=\alpha _{R}$), the 1D equal Rashba and Dresselhaus SO coupling
can be generated. Fortunately, this 1D SO coupling has been realized
experimentally in the ultracold neutral atoms \cite{Lin,PW,LWC}.

In the absence of SO coupling, the limit $\mathbf{q=0}$ can be applied
usually to discuss the standard two-channel model including the effective
Zeeman field \cite{DES}. However, in the presence of SO coupling, the result
is quite complicated. If both SO coupling and the effective Zeeman field are
taken into account, the parity and time-reversal symmetries are broken. As a
result, the $\mathbf{q}$-dependent order parameter should be introduced \cite%
{DFA}. However, in this paper we do not consider the effect of the effective
Zeeman field and thus may focus on the case of $\mathbf{q=0}$.

\section{Two-body bound state}

We begin to discuss the two-body bound state of the generalized two-channel
model (\ref{TOTAL H}) by introducing the ansatz wavefunction. In the absence
of SO coupling ($\alpha =0$), Hamiltonian (\ref{TOTAL H}) reduces to the
standard two-channel model $H=\sum_{\mathbf{k}\sigma }\xi _{\mathbf{k}}C_{%
\mathbf{k}\sigma }^{\dagger }C_{\mathbf{k}\sigma }+\sum_{{}}(\delta
_{0}-2\mu )b_{0}^{\dagger }b_{0}+g\sum_{\mathbf{kk}^{\prime }}b_{0}^{\dagger
}C_{-\mathbf{k}\downarrow }C_{\mathbf{k}\uparrow }+C_{\mathbf{k}^{\prime
}\uparrow }^{\dagger }C_{-\mathbf{k}^{\prime }\downarrow }^{\dagger }b_{0}$
\cite{MH,ETI,YO,RDU}, in which only the singlet Cooper paring can be formed.
However, in the presence of SO coupling, both the singlet and triplet Cooper
parings can coexist \cite{LPG}. As a result, the ansatz wavefunction should
be written formally as
\begin{equation}
\left\vert \Psi \right\rangle =(\sum_{k^{^{\prime }}\sigma \sigma ^{^{\prime
}}}\beta _{k^{^{\prime }}\sigma \sigma ^{^{\prime }}}C_{\mathbf{k}^{\prime
}\sigma }^{\dagger }C_{\mathbf{k}^{\prime }\sigma ^{^{\prime }}}^{\dagger
}+\gamma b_{0}^{\dagger })\left\vert 0,0,0,0\right\rangle \otimes \left\vert
0\right\rangle ,  \label{WS}
\end{equation}%
where $\beta _{k^{^{\prime }}\uparrow \downarrow }$ and $\beta _{k^{^{\prime
}}\downarrow \uparrow }$ ($\beta _{k^{^{\prime }}\uparrow \uparrow }$ and $%
\beta _{k^{^{\prime }}\downarrow \downarrow }$) represent the probability
amplitude of the singlet (triplet) Cooper paring, $\gamma $ stands for the \
probability amplitude of the molecule, and $\left\vert 0,0,0,0\right\rangle
\otimes \left\vert 0\right\rangle $ is the direct multiple of the fermion
vacuum with spin flipping and the molecule vacuum. Substituting the
wavefunction in Eq. (\ref{WS}) into the stationary Schr\"{o}dinger equation
\begin{equation}
(H-E)\left\vert \Psi \right\rangle =0,  \label{SS}
\end{equation}%
we find that the six coefficients determining the ansatz wavefunction $%
\left\vert \Psi \right\rangle $ and the energy $E$ are governed by the
following equations:
\begin{equation}
\left\{
\begin{array}{c}
g\gamma +\beta _{k^{^{\prime }}\uparrow \uparrow }\alpha k_{-}=\Xi _{k}\beta
_{k\downarrow \uparrow } \\
g\gamma +\beta _{k^{^{\prime }}\downarrow \downarrow }\alpha k_{+}=\Xi
_{k}\beta _{k\uparrow \downarrow } \\
(\beta _{k^{^{\prime }}\downarrow \uparrow }+\beta _{k^{^{\prime }}\uparrow
\downarrow })\alpha k_{+}=\Xi _{k}\beta _{k^{^{\prime }}\uparrow \uparrow }
\\
(\beta _{k^{^{\prime }}\downarrow \uparrow }+\beta _{k^{^{\prime }}\uparrow
\downarrow })\alpha k_{-}=\Xi _{k}\beta _{k^{^{\prime }}\downarrow
\downarrow } \\
2(\delta _{0}-E)\gamma =-g\sum_{k^{^{\prime }}}(\beta _{k^{^{\prime
}}\uparrow \downarrow }+\beta _{k^{^{\prime }}\downarrow \uparrow })%
\end{array}%
\right. ,  \label{EQC}
\end{equation}%
where $\Xi _{k}=E-2\epsilon _{k}$ and $k_{\pm }=k_{y}\pm i\lambda k_{x}$.
Eq. (\ref{EQC}) can not be solved directly because of lack of a coefficient
equation. If we define the spin symmetry and anti-symmetry vectors as
\begin{equation}
\left\{
\begin{array}{c}
\psi _{s}(k)=\frac{1}{\sqrt{2}}(\beta _{k\downarrow \uparrow }-\beta
_{k\uparrow \downarrow }) \\
\psi _{a}(k)=\frac{1}{\sqrt{2}}(\beta _{k\downarrow \uparrow }+\beta
_{k\uparrow \downarrow })%
\end{array}%
\right. ,  \label{Vect}
\end{equation}%
the stationary Schr\"{o}dinger equation is rewritten as $(H-E)\psi _{k}=0$
in the representation of $\psi _{k}=[\beta _{k\uparrow \downarrow },\beta
_{k\downarrow \uparrow },\beta _{k\downarrow \downarrow },\beta _{k\uparrow
\uparrow },\gamma ]^{T}$. This leads to another equations for the
coefficients $\beta _{k^{^{\prime }}\sigma ,\sigma ^{^{\prime }}}$ and $%
\gamma $, that is,
\begin{equation}
\left\{
\begin{array}{c}
\gamma =-\sum_{k^{^{\prime }}}\frac{g\sqrt{2}\psi _{a}(k^{^{\prime }})}{%
(\delta _{0}-E)} \\
\beta _{k^{^{\prime }}\uparrow \uparrow }=\frac{\sqrt{2}\psi _{a}(k)\alpha
k_{+}}{\Xi _{k}} \\
\beta _{k^{^{\prime }}\downarrow \downarrow }=\frac{\sqrt{2}\psi
_{a}(k)\alpha k_{-}}{\Xi _{k}}%
\end{array}%
\right. .  \label{ACE}
\end{equation}%
Substituting Eq. (\ref{ACE}) into Eq. (\ref{EQC}) yields
\begin{equation}
\sum_{k}(\frac{\Xi _{k}}{\Xi _{k}^{2}-4\alpha ^{2}k_{+}k_{-}}+\frac{1}{%
2\epsilon _{k}})=\frac{E-\delta }{g^{2}}.  \label{MEQ}
\end{equation}

\begin{figure}[t]
\includegraphics[width=7.5cm]{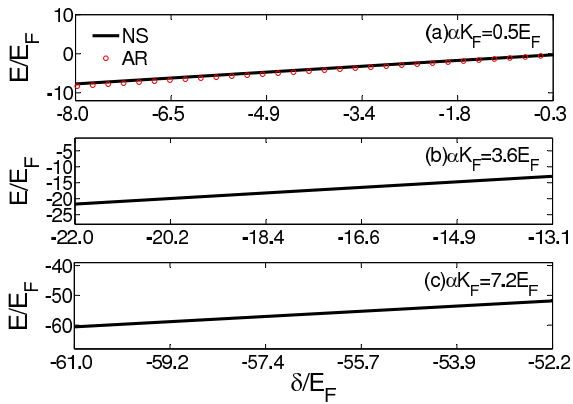}\newline
\caption{(Color online) The bound-state energy for the 2D Rashba SO coupling
($\protect\lambda =1$) as a function of the detuning $\protect\delta $ for
the different strengths of SO coupling (a) $\protect\alpha K_{\text{F}%
}=0.5E_{\text{F}}$, (b) $\protect\alpha K_{\text{F}}=3.6E_{\text{F}}$, and
(c) $\protect\alpha K_{\text{F}}=7.2E_{\text{F}}$ when $\Gamma =0.1$. In
Fig. 1(a), the red open symbol corresponds to the analytical result (AR) and
the black solid line represents the direct numerical simulation (NS).}
\label{fig1}
\end{figure}

Equation (\ref{MEQ}), which is the main result of this paper, determines the
SO-induced bound-state energy of the generalized two-channel model (\ref%
{TOTAL H}). The procedure is given as follows. (i) We first obtain the
energy $E$ from Eq. (\ref{MEQ}). (ii) Then, we introduce the threshold
energy $E_{T}$, which is the lowest energy of the free particle (i.e., the
lowest band), to judge whether this energy is the bound-state energy or not.
If $E<E_{T}$, the bound state exists and the corresponding energy $E$ is
called the bound-state energy, and vice versa \cite{JPV}. According to its
definition, the threshold energy $E_{T}$ of the generalized two-channel
model (\ref{TOTAL H}) is evaluated as
\begin{equation}
E_{T}=-\frac{2m\alpha ^{2}}{\hbar ^{2}}.  \label{ET}
\end{equation}%
In the absence of SO coupling ($\alpha =0$), this threshold energy becomes $%
E_{T}=0$, as expected. In the following discussions, we mainly consider two
interesting cases, including the 2D Rashba SO coupling ($\lambda =1$) and 1D
equal Rashba and Dresselhaus SO coupling ($\lambda =0$), to reveal the
fundamental properties of the bound state.

We first address the case of the 2D Rashba SO coupling ($\lambda =1$). In
the absence of SO coupling ($\alpha =0$), the analytical bound-state energy
is derived from Eq. (\ref{MEQ}) by
\begin{equation}
E=\frac{1}{32\pi ^{2}\hbar ^{6}}(-g^{4}m^{3}+32\pi ^{2}\delta \hbar
^{6}-g^{2}\sqrt{g^{4}m^{6}-64m^{3}\pi ^{2}\delta \hbar ^{6}}).  \label{EN}
\end{equation}%
It implies that in such a case only one bound state can be found \cite{DES}.
In the presence of SO coupling ($\alpha \neq 0$), the explicit expression
for the bound-state energy can not be obtained. However, for the weak SO
coupling, Eq. (\ref{MEQ}) is simplified as
\begin{equation}
\frac{m^{\frac{3}{2}}[2\hbar ^{2}E+(1+\lambda ^{2})m\alpha ^{2}]}{8\pi \hbar
^{5}\sqrt{-E}}=\frac{E-\delta }{g^{2}}  \label{EEQ}
\end{equation}%
with the help of a Taylor expansion with respect to the strength of SO
coupling. In this case, Hamiltonian (\ref{TOTAL H}) also exhibits one bound
state, as shown in Fig. 1(a). By further solving Eq. (\ref{EEQ})
approximately, we find that the bound-state energy is proportional to $%
-\alpha ^{4}$. This behavior agrees well with the numerical simulation, as
also shown in Fig. 1(a). It implies that the bound-state energy can decrease
by increasing the strength of SO coupling. For the strong SO coupling, the
perturbation method is invalid. Here we numerically solve Eq. (\ref{MEQ}) to
evaluate the bound-state energy $E$. Even if the strong SO coupling exists,
only one bound state can be found, as shown in Figs. 1(b) and 1(c)
\begin{figure}[t]
\includegraphics[width=7.4cm]{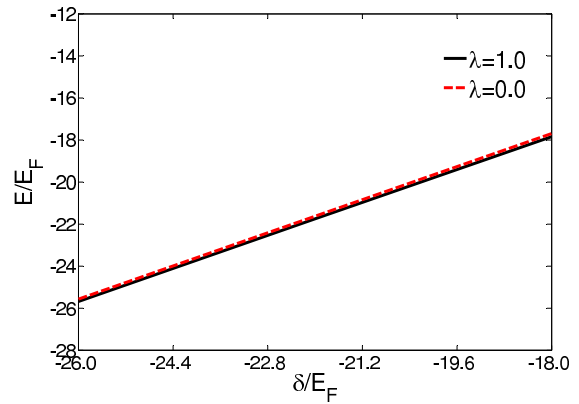}\newline
\caption{(Color online) The bound-state energy as a function of the detuning
$\protect\delta $ for the different types of SO coupling including $\protect%
\lambda =1$ (the 2D Rashba SO coupling) and $\protect\lambda =0$ (the 1D
equal Rashba and Dresselhaus SO coupling) when $\Gamma =0.1$ and $\protect%
\alpha K_{\text{F}}=4.2E_{\text{F}}$.}
\label{fig2}
\end{figure}

In Fig. \ref{fig2}, we plot the bound-state energy with respect to the
detuning $\delta $ for the different types of SO coupling including $\lambda
=1$ (the 2D Rashba SO coupling) and $\lambda =0$ (the 1D equal Rashba and
Dresselhaus SO coupling). It can be seen clearly that the bound-state energy
is affected slightly by the type of SO coupling.

\begin{figure}[tbp]
\includegraphics[width=7.9cm]{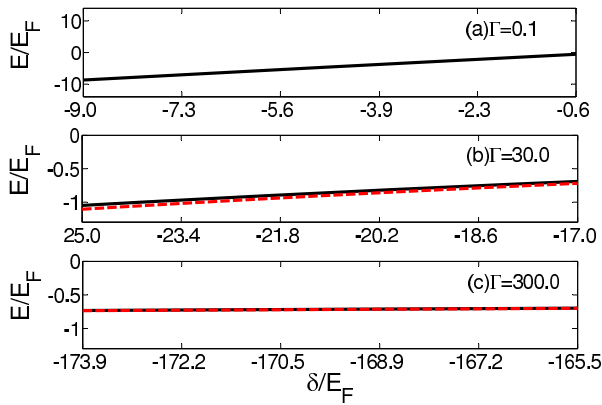}\newline
\caption{(Color online) The bound-state energy for the 1D equal Rashba and
Dresselhaus SO coupling with $\protect\lambda =0$ as a function of the
detuning $\protect\delta $ for the different Feshbach-resonant widths (a) $%
\Gamma =0.1$, (b) $\Gamma =30$, and (c) $\Gamma =300.0$ when $\protect\alpha %
K_{F}=0.7E_{F}$. In (b) and (c), the red dash line corresponds to the
numerical solution of Eq. (\protect\ref{EEQN}). For the 2D Rashba SO
coupling, the similar conclusion can be also found.}
\label{fig3}
\end{figure}

When the Feshbach-resonant width parameter $\Gamma $ increases, the system
changes from the narrow limit ($\Gamma \ll 1$) to the broad limit ($\Gamma
\gg 1$). Especially, for the broad limit ($\Gamma \gg 1$), Eq. (\ref{MEQ})
becomes
\begin{equation}
\sum_{k}(\frac{\Xi _{k}}{\Xi _{k}^{2}-4\alpha ^{2}k_{+}k_{-}}+\frac{1}{%
2\epsilon _{k}})\simeq -\frac{\delta }{g^{2}}=\frac{m}{4\pi \hbar ^{2}a_{s}},
\label{EEQN}
\end{equation}%
which is similar to the result of Ref. \cite{YZQ}. In Fig. \ref{fig3}, we
plot the bound-state energy for the 1D equal Rashba and Dresselhaus SO
coupling as a function of the detuning $\delta $ for the different
Feshbach-resonant width parameters (a) $\Gamma =0.1$, (b) $\Gamma =30.0$,
and (c) $\Gamma =300.0$. This figure shows again that for the broad limit
our considered two-channel model reduces to the single-channel model.
However, it should be remarked that the energy $E$ for the broad Feshbach
resonance is not a bound-state energy, but is a two-body interaction energy
approaching infinitely the bound-state energy \cite{LDL}

\section{Molecular signature}

Having obtained the bound-state energy in the generalized two-channel model,
it is conveniently to discuss the experimentally-measurable molecular
signature. In terms of the Hellmann-Feymann theorem, the molecular fraction
is obtained by
\begin{equation}
N_{0}=\left\langle \Psi \right\vert b_{0}^{\dagger }b_{0}\left\vert \Psi
\right\rangle =\left\langle \Psi \right\vert \frac{\partial H}{\partial
\delta _{0}}\left\vert \Psi \right\rangle =\frac{dE}{d\delta },  \label{MN}
\end{equation}%
where the bound-state energy $E$ can be derived from Eq. (\ref{MEQ}).

\begin{figure}[tbp]
\includegraphics[width=8cm]{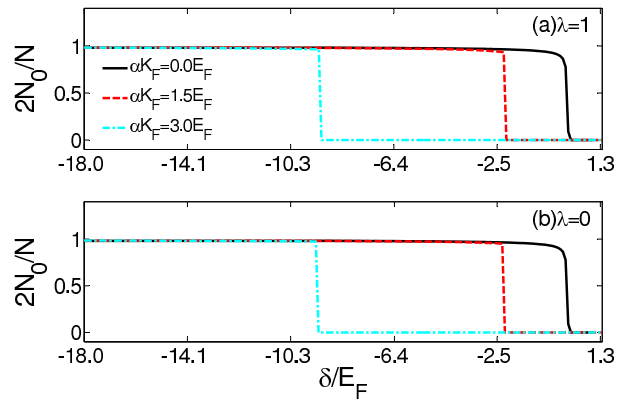}\newline
\caption{(Color online) The molecular fraction of both (a) the 2D Rashba SO
coupling with $\protect\lambda =1$ and (b) the 1D equal Rashba and
Dresselhaus SO coupling with $\protect\lambda =0$ as a function of the
detuning $\protect\delta $ for the different strengths of SO coupling when $%
\Gamma =0.1$.}
\label{fig4}
\end{figure}

Figure \ref{fig4} is plotted the scaled molecular fraction $2N_{0}/N$ of
both the 2D Rashba SO coupling ($\lambda =1$) and the 1D equal Rashba and
Dresselhaus SO coupling ($\lambda =0$) with respect to the detuning $\delta $
for the different strengths of SO coupling. In the absence of SO coupling ($%
\alpha =0$), the molecule exists in the negative detuning ($\delta <0$).
However, for the positive detuning ($\delta \geq 0$), the physical bound
state vanishes \cite{LDL}, i.e., there is no real molecular fraction. With
the increasing of the strength of SO coupling, the critical point at which
the molecular fraction vanishes shifts from zero to the negative detuning.
The physics can be understood as follows. In the generalized two-channel
model, the molecules play two roles. One is that they interact directly with
the atoms via Hamiltonian $H_{\text{I}}$. The other (the most important) is
that they induce the indirect atom-atom interactions, which generate the
Cooper pairing. When the Cooper pairing is enhanced by SO coupling \cite%
{GM,YZQ,HH}, the molecules are thus suppressed because the system need
guarantee a conserved number $N=2b_{0}^{\dagger }b_{0}+\sum_{\mathbf{k}%
\sigma }C_{\mathbf{k}\sigma }^{\dagger }C_{\mathbf{k}\sigma }$. In order to
see clearly this behavior induced by SO coupling, we introduce a key
parameter $\delta _{m}$. This parameter describes the maximum detuning at
which the molecular fraction exists. In terms of the definition, the
parameter $\delta _{m}$ is given by%
\begin{equation}
\delta _{m}=-\frac{2m\alpha ^{2}}{\hbar ^{2}}-g^{2}[\sum_{k}(\frac{\Xi _{k}}{%
\Xi _{k}^{2}-4\alpha ^{2}k_{+}k_{-}}+\frac{1}{2\epsilon _{k}})]_{E=E_{T}}.
\label{DM}
\end{equation}%
In the case of the weak SO coupling, the parameter $\delta _{m}$ is obtained
explicitly by%
\begin{equation}
\delta _{m}=-\frac{32\hbar ^{2}m\pi \alpha ^{2}+\sqrt{2}g^{2}m^{2}(\lambda
^{2}-3)\alpha }{16\pi \hbar ^{4}}\simeq -\alpha ^{2}.  \label{DMEQ}
\end{equation}%
Eq. (\ref{DMEQ}) shows clearly that $\delta _{m}$ decreases with the
increasing of the strength of SO coupling.

\begin{figure}[t]
\includegraphics[width=8cm]{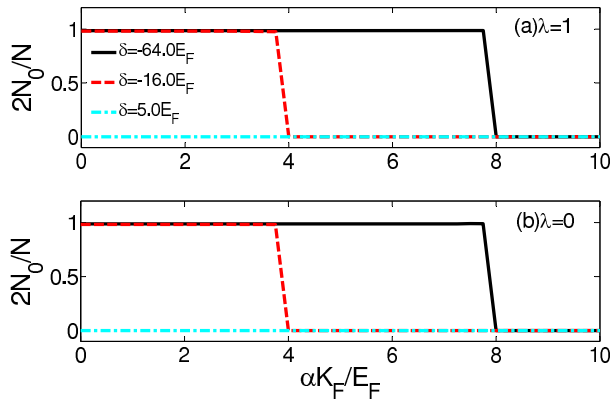}\newline
\caption{(Color online) The molecular fraction of both (a) the 2D Rashba SO
coupling with $\protect\lambda =1$ and (b) the 1D equal Rashba and
Dresselhaus SO coupling with $\protect\lambda =0$ as a function of the
strength $\protect\alpha K_{\text{F}}$ of SO coupling for the different
detunings when $\Gamma =0.1$.}
\label{fig5}
\end{figure}

In Fig. \ref{fig5}, we plot the molecular fraction of both the 2D Rashba SO
coupling ($\lambda =1$) and the 1D equal Rashba and Dresselhaus SO coupling (%
$\lambda =0$) with respect to the strength of SO coupling for the different
detunings. In the negative detuning ($\delta <\delta _{m}$) we shows again
that the SO coupling suppresses the molecular fraction. With the increasing
of the strength of SO coupling, the molecular fraction also vanishes. It means
that in experiment the molecular fraction can be well controlled by tuning
the SO strength. In addition, in the positive detuning, no molecular
fraction can be found even if SO coupling exists.

\section{Conclusions and Remarks}

In summary, motivated by the recent experimental developments, we have
investigated the SO-driven degenerate Fermi gas in the narrow Feshbash
resonance based on the generalized two-channel model. We have found that in
the absence of the atom-atom interactions, only one bound state can be found
even if SO coupling exists. In addition, we have shown that the molecular
fraction can be well controlled by SO coupling. We believe that in
experiments it is a good signature to explore the SO-induced physics.

\section{Acknowledgements}

We thank Professors Peng Zhang, Wei Yi, Wei Zhang, Shizhong Zhang and Doctor
Zengqiang Yu for their helpful discussions. This work was supported partly
by the 973 program under Grant No. 2012CB921603; the NNSFC under Grants No.
10934004, No. 11074154, and No. 61275211; NNSFC Project for Excellent
Research Team under Grant No. 61121064; and International Science and
Technology Cooperation Program of China under Grant No.2001DFA12490.

\emph{Note added}--During preparing this paper, we noticed that two bound
states for the SO-driven two-channel model with the atom-atom interactions
was predicted by V. B. Shenoy in terms of a renormalizable quantum field
theory \cite{VBS1}. However, that paper does not consider the molecular
fraction.

\end{document}